\documentclass[pre preprint]{revtex4}

\def\be{\begin{equation}}
\def\ee{\end{equation}}
\def\ba{\begin{eqnarray}}
\def\ea{\end{eqnarray}}

\begin{document}
\title{Application of Dressing Method for Long Wave-Short Wave
Resonance Interaction Equation}

\author{Maria Yurova}

\email{yurova-m@rambler.ru}

\affiliation{Institute of Nuclear Physics\\Moscow State University
\\Vorobjovy Gory\\119899 Moscow\\Russia}


\begin{abstract}
\noindent In this paper we investigate the application of Zakharov -
Shabat dressing method to (2+1) - dimensional long wave - short wave
resonance interaction equation (LSRI). Using this method we can
construct the exact $N$ - soliton solution of this equation
depending on arbitrary constants. It contains both solutions which
don't decay along $N$ different directions in space infinity, and
"dromion" ones, or localized solutions that decay exponentially in
all directions.
\end{abstract}

\pacs{02.40Xx,47.35Fg}

\maketitle


\section{Introduction}

We investigate  the  (2+1) - dimensional equation describing the
interaction of a long wave with a packet of short waves. Such
process can arise in fluid mechanics. If $L$ is a long interfacial
wave and $S$ is a short waves packet, their interaction on
$(x,y)$-plane is described by the system \cite{Mel'nikov} \ba  i
\partial _{t} S & + & i c_{g}\partial _{x} S - \beta L S -
\gamma
\partial
^{2}_{xx}S - \delta |S|^{2} S = 0,\nonumber\\
 \partial _{t} L & + & c_{l}\partial _{y} L + \alpha \partial _{x} |S|^{2}  =  0,\nonumber
\ea
 where $c_{l}$ is a long wave phase velocity along the $y$-axis,
$c_{g}$ is a group velocity of a packet of short waves along the
$x$-axis, and $\alpha$, $\beta$, $\gamma$, $\delta$ are constant
parameters of a system under consideration. Using the "resonance
condition" $c_{l} = c_{g}$ \cite{Dodd}, some linear coordinate
transformations and scale transformation of the constants, it is
possible to rewrite this system in the form: \ba
&i \partial _{t} S & - \,\, \partial ^{2}_{xx} S + L\,S = 0,\\
&\partial _{y} L & =  2 \partial _{x} |S|^{2}\nonumber. \ea The
integrability of this system has been established earlier.
Therefore, a Lax pare for equation (1)has been constructed in
\cite{ZMNP}, and an exact solution for this equation has been
presented in \cite{Mel'nikov}. The LSRI equation considered in
\cite{LC} and \cite{RKLTL} can be written down in our form by simple
linear coordinate transformations. In \cite{LC} the authors have
performed a deep analysis of different classes of exact solutions
such as "positon", "dromion" and "solitoff", and they have derived a
number of new solutions including one on a continuous wave
background. The recent investigation of the above system has been
performed in \cite{RKLTL}. The authors showed that (1) possess the
Painleve property and they generated an extended class of periodic
wave solutions and exponentially localized ones.

In this paper we consider the application of Zakharov - Shabat
dressing method \cite{ZS} to the above system. Using the general
scheme we obtain the exact $N$ - soliton solution which depends on
arbitrary constants. The general form of this solution is new. The
analysis of asymptotic behaviour demonstrates that this solution
describes an interaction of $N$ solitons and does not decay along
$N$ different directions in space infinity. For $N = 1$ the
asymptotic form of the solution obtained coincides with the solution
of \cite{Mel'nikov}. If we impose some restrictions on the constants
we obtain the "dromion" i.e. localized solutions, that decay
exponentially in all directions.


\section{The dressing method}
Now, let us reformulate the system under consideration in the
coordinates $\xi$ and $\eta$, where \be
\partial_{\xi} = \partial_{x} - \partial_{y},\,\,\,\partial_{\eta} =
\partial_{x} \ee to apply the general inverse scattering problem
method for integration of nonlinear differential equations, proposed
by Zakharov and Shabat \cite{ZS}. Then equation (1) has the form \ba
i \partial _{t} S - \partial ^{2}_{\eta\eta} S + L\,S = 0,\nonumber\\
(\partial _{\eta} - \partial _{\xi})\,L = 2 \partial _{\eta}
|S|^{2}. \ea Following the above method \cite{ZS}, we consider a
linear integral operator $\widehat{F}$, acting to function $ \phi$
\be \widehat{F} \phi (\xi,\eta,t) = \int _{-\infty}^{\infty}
F(\xi,z,\eta,t)\phi (z,\eta,t) dz,\nonumber \ee where the matrix $F$
satisfy the condition \be \sup_{\xi > \xi_{0}} \int _{\xi_{0}}
^{\infty} |F(\xi,z,\eta,t)|dz < \infty \nonumber\ee for all $\xi_{0}
> -\infty$. Let the operator $\widehat{F}$ have the
form \be \widehat{1} + \widehat{F} = (\widehat{1} +
\widehat{K}_{+})^{-1}(\widehat{1} + \widehat{K}_{-}), \ee where
$\widehat{K}_{+}$ and $\widehat{K}_{-}$ are Volterra operators \be
\pm \widehat{K}_{\pm} \phi (\xi,\eta,t) = \int _{\xi}^{\pm\infty}
K_{\pm}(\xi,z,\eta,t)\phi (z,\eta,t) dz,\nonumber \ee and
$K_{+}(\xi,z,\eta,t) = 0$ for $z < \xi$, $K_{-}(\xi,z,\eta,t) = 0$
for $z > \xi$. The kernel $K_{+}$ satisfy the
Gelfand-Levitan-Marchenko equation \be F(\xi,z) + K_{+}(\xi,z) +
\int_{\xi} ^{\infty} K_{+}(\xi,s) F(s,z) ds = 0 \,\, (z>\xi). \ee
Now let us define two primeval operators, acting to function $ \phi$
in the form: \ba \widehat{L^{(1)}_{0}} & = &
\partial_{t} - i \,\,\,\,
\sigma^{+} \,\partial^{2}_{\xi\xi},\nonumber\\
\widehat{L^{(2)}_{0}} & = & \partial_{\eta} - \,\,
\sigma^{+} \,\partial_{\xi}, \,\,\sigma^{+} = \left (\begin{array}{crc} 1 & 0 \\ 0 & 0 \\
\end{array}\right ),
\end{eqnarray}
and consider a class of differential operators $\widehat{L^{(1)}}$,
$\widehat{L^{(2)}}$, obtained from $\widehat{L^{(1)}_{0}}$ and
$\widehat{L^{(2)}_{0}}$ by the transformation: \be
\widehat{L^{(i)}}\,\,(\widehat{1} + \widehat{K_{+}}) = (\widehat{1}
+ \widehat{K_{+}}) \,\,\widehat{L^{(i)}_{0}},\,\,\, i=1,2, \ee where
$\widehat{1} + \widehat{K_{+}}$ is Volterra operator. Since the
differential parts of operator relation (10) are equal to zero,
$\widehat{L^{(1)}}$ and $\widehat{L^{(2)}}$ have the form: \ba
\widehat{L^{(1)}} = \widehat{L^{(1)}_{0}} + u_{1}^{(1)} \partial
_{\xi} + u_{0}^{(1)}, \,\,\widehat{L^{(2)}} = \widehat{L^{(2)}_{0}}
+ u_{0}^{(2)}. \ea Equation (10) gives the reccurent relations
between the matrix coefficients $u_{k}^{(i)}$, $i=1,2$, and
components of matrix kernel $K_{+}(\xi,z,\eta,t)$. It turned out,
that if the components of $K_{+}$ satisfy some conditions, we can
obtain the Lax pare for the system (1). Let \ba
K_{+}(\xi,z,\eta,t) = \left (\begin{array}{crc} k_{11} & k_{12} \\ k_{21} & k_{22} \\
\end{array}\right ), \,\,k_{ij} = k_{ij}(\xi,z,\eta,t). \ea
Then, if we impose the following conditions: \ba (\partial_{\eta} -
\partial_{\xi} - k_{22})(k_{12}|_{z=\xi}) +
\partial_{z}k_{12}|_{z=\xi}=0,\nonumber\\
(\partial_{\eta} + k_{11})(k_{21}|_{z=\xi})
-\partial_{z}k_{21}|_{z=\xi}=0,\nonumber\\
k_{12}|_{z=\xi} = - \bar{k}_{21}|_{z=\xi},\,\,k_{11}|_{z=\xi} =
\bar{k}_{11}|_{z=\xi}, \ea we can identify a short waves packet $S$
and a long interfacial wave $L$ with $k_{ij}$ as follow: \ba S&= &
k_{21}(\xi,z,\eta,t)|_{z=\xi},\nonumber\\ L & = & 2\,[S\bar S -
\partial _{\xi} (k_{11}(\xi,z,\eta,t)|_{z=\xi})]. \ea Thus, the
matrix coefficients $u_{k}^{(i)}$ can be expressed as the functions
of $S$ and $L$: \ba
u_{0}^{(1)} &=& i\left (\begin{array}{crc} -S\bar S + L &  (\partial _{\eta} + \partial _{\xi})\bar S \\
-\partial _{\eta}S & -S\bar S \\
\end{array}\right ),\nonumber\\
- i u_{1}^{(1)} &=& u_{0}^{(2)} = \left (\begin{array}{crc} 0 & \bar S \\ S & 0 \\
\end{array}\right ).
\ea Consequently, it is possible to demonstrate by straightforward
calculation, that the integrability condition for the "dressing"
operators (11) leads to equation (1) under consideration, and we
obtain two commuting operators \be
[\widehat{L^{(1)}},\widehat{L^{(2)}}] = 0, \ee that constitute the
Lax pare for equation (1). The same form of Lax pare has been
obtained in \cite{ZMNP}.


\section{Exact Solution}
Now, let us construct the exact solution of LSRI equation using the
above method. Two commutation relations for operator $\widehat{F}$
\be [\widehat{L^{(1)}_{0}}, \widehat{F}]
=0,\,\,\,[\widehat{L^{(2)}_{0}}, \widehat{F}] =0, \ee give two
differential equations for the kernel $F$: \ba
 &\partial _{t}F & - \,\,i \,\,
\sigma^{+} \, {\partial _{\xi}}^{2} F - i\,\, \partial _{z}^{2}F
\,\, \sigma^{+} \, = 0,\nonumber\\
& \partial _{\eta}F & - \,\, \sigma^{+} \, \partial _{\xi} F -
\partial _{z} F
\,\, \sigma^{+} \, = 0. \ea We choose the solution of these
equations in the form: \ba
F & = & \sum_{n}^{N}\left (\begin{array}{crc} f^{n}_{11} & f^{n}_{12} \\ f^{n}_{21} & f^{n}_{22} \\
\end{array}\right ), \nonumber\\
f^{n}_{11} & = & {(f^{n}_{11})}_{0}\, e^{-\kappa^{n}_{1}
(2\eta+\xi+z)},
\nonumber\\
f^{n}_{12} & = & {(f^{n}_{12})}_{0}\, e^{i(\kappa^{n}_{2})^{2}t
-\kappa^{n}_{2}(\eta + \xi) -
\bar{\kappa}^{n}_{3}z},\nonumber\\
f^{n}_{21} & = & -{(\bar{f}^{n}_{12})}_{0}\,
e^{-i(\bar{\kappa}^{n}_{2})^{2}t -\bar{\kappa}^{n}_{2}(\eta + z)
- \kappa^{n}_{3} \xi} ,\nonumber\\
f^{n}_{22} & = & {(f^{n}_{22})}_{0}\, e^{-\kappa^{n}_{4}
(\xi+z)},\ea where \ba \Re {(f^{n}_{ij})}_{0}  > 0,\,\, i,j
=1,2,\,\,\Im {(f^{n}_{11})}_{0}  = \Im {(f^{n}_{22})}_{0}
 = 0 , \nonumber \ea and $ \kappa^{n}_{i}$ = $\bar
{\kappa}^{n}_{i}$, $i=1,4$, $\kappa^{n}_{i}$ = $\Re \kappa^{n}_{i} +
i\Im \kappa^{n}_{i}$, $i=2,3$, $\Re \kappa_{i} > 0$, $i=1,..4$ are
constant parameters. Solving the equation (8), we obtain the
explicit form of
 the matrix kernel $K_{+}(\xi,z,\eta,t)$, which is
too complicated and we omit it here. Further using the expressions
(14) we can write down the exact $N$ - solitons solution for the
short waves packet $S$ and a long interfacial wave $L$. Returning to
usual space - time variables $x, y, t$, we obtain the explicit form
of our solution:
 \ba S & = & \sum_{n}^{N}\,\frac{1}{\det A}\,\,
 \left |\begin{array}{crc} A & B^{n} \\ \alpha & \beta^{n} \\
\end{array}\right |, \nonumber\\
L & = & 2\,\,[S \bar S \, -\, (\partial_{x} -
\partial_{y})\sum_{n}^{N}\,\frac{1}{\det A}\,\,
 \left |\begin{array}{crc} A & B^{n} \\ \tilde{\alpha} & \tilde{\beta^{n}} \\
\end{array}\right |], \ea where the block $4N \times 4N$ matrix $A$, the $4N \times 1$
column $B^{n}$, the $1 \times 4N$ rows $\alpha $ and
$\tilde{\alpha}$ and the scalars $\beta^{n}$ and $\tilde{\beta^{n}}$
have the form:
\ba A &  = &\left (\begin{array}{cccc} I_{N} + a_{11} & a_{12} & 0 & 0 \\
a_{21} & a_{22} & I_{N} & 0 \\ 0 & I_{N} & a_{33} & a_{34} \\
0 & 0 & a_{43} & I_{N} + a_{44} \\ \end{array}\right ),\nonumber\\
B^{n} &=& \left(
  \begin{array}{c}
    e^{\kappa^{n}_{1}y} (a_{11})_{in} \\
    e^{\kappa^{n}_{1}y} (a_{21})_{in} \\
    e^{\bar{\kappa}^{n}_{2}y} (a_{33})_{in} \\
    e^{\bar{\kappa}^{n}_{2}y} (a_{43})_{in} \\
  \end{array}
\right), \nonumber\\
 \alpha & = & \left(
  \begin{array}{cccc}
    0 & 0 & e^{-\bar {\kappa}^{n}_{2}x -
(\bar {\kappa}^{n}_{2} - \kappa^{n}_{3})y} {(\bar
{f}^{n}_{12})}_{0}e^{ - i (\bar {\kappa}^{n}_{2})^{2}t} &
e^{\kappa^{n}_{4}y} {(f^{n}_{22})}_{0}  \\
  \end{array}
\right),\nonumber\\  \tilde{\alpha} & = & \left(
  \begin{array}{cccc}
   e^{-2\kappa^{n}_{1}x -
\kappa^{n}_{1}y} \, {(f^{n}_{11})}_{0}  & - e^{- \kappa^{n}_{2}x}\,
{(f^{n}_{12})}_{0}\,e^{ i ({\kappa}^{n}_{2})^{2}t}
 & 0 & 0 \\
  \end{array}
\right),\nonumber\\ \beta^{n} & = &
e^{\bar{\kappa}^{n}_{2}y}\,{\alpha}_{n},\,\, \tilde{\beta^{n}}  =
e^{{\kappa}^{n}_{1}y}\,{\tilde{\alpha}}_{n}; \ea $I_{N}$ is $N
\times N$ unit matrix and $0$ are $N \times N$ or $1 \times N$
matrices respectively, $a_{IJ} \equiv \|(a_{IJ})_{ij}\|$ are $N
\times N$ matrices with components: \ba (a_{11})_{ij}& = &
\frac{e^{(\kappa^{i}_{1} - \kappa^{j}_{1})y - 2\kappa^{j}_{1}x}
{(f^{j}_{11})}_{0}}{\kappa^{i}_{1} + \kappa^{j}_{1}},\nonumber\\
(a_{21})_{ij} & = & \frac{e^{(\bar{\kappa}^{i}_{2} -
\kappa^{j}_{1})y - 2\kappa^{j}_{1}x}
{(f^{j}_{11})}_{0}}{\bar{\kappa}^{i}_{2} +
\kappa^{j}_{1}},\nonumber\\(a_{12})_{ij} &=& \frac{-
e^{\kappa^{i}_{1}y - {\kappa}^{j}_{2}x}
{(f^{j}_{12})}_{0}}{\kappa^{i}_{1} + {\kappa}^{j}_{2}}\,e^{ i
({\kappa}^{j}_{2})^{2}t}, \nonumber\\ (a_{22})_{ij} &=& \frac{-
e^{\bar{\kappa}^{i}_{2}y - {\kappa}^{j}_{2}x}
{(f^{j}_{12})}_{0}}{\bar{\kappa}^{i}_{2} +
{\kappa}^{j}_{2}}\,e^{ i ({\kappa}^{j}_{2})^{2}t}, \nonumber\\
(a_{33})_{ij} & = & \frac{e^{(\bar{\kappa}^{i}_{3} + \kappa^{j}_{3}
- \bar{\kappa}^{j}_{2})y - \bar{\kappa}^{j}_{2}x}
{(\bar{f}^{j}_{12})}_{0}}{\bar{\kappa}^{i}_{3} +
\kappa^{j}_{3}}\,e^{- i (\bar{\kappa}^{j}_{2})^{2}t},\nonumber\\
(a_{43})_{ij} & = & \frac{e^{(\kappa^{i}_{4} + \kappa^{j}_{3} -
\bar{\kappa}^{j}_{2})y - \bar{\kappa}^{j}_{2}x}
{(\bar{f}^{j}_{12})}_{0}}{\kappa^{i}_{4} + \kappa^{j}_{3}}\,e^{- i
(\bar{\kappa}^{j}_{2})^{2}t},\\(a_{34})_{ij} &=& \frac{-
e^{(\bar{\kappa}^{i}_{3} + \kappa^{j}_{4})y}
 {(f^{j}_{22})}_{0}}{\bar{\kappa}^{i}_{3} +
\kappa^{j}_{4}},\,\, (a_{44})_{ij} = \frac{ e^{(\kappa^{i}_{4} +
\kappa^{j}_{4})y } {(f^{j}_{22})}_{0}}{\kappa^{i}_{4} +
\kappa^{j}_{4}}.\nonumber \ea In general case this solution
describes the intersection of $N$ solitons. The analysis of its
asymptotic behaviour demonstrates that it vanishes at spatial
infinity $x, y \rightarrow \infty$ in all directions except the $N$
ones \be \frac{y}{x} = \frac{\Re\kappa^{i}_{2}}{\Re\kappa^{i}_{3}},
i = 1,...N.\ee  In the simplest case of $N = 1$ the one - soliton
solution for the short waves packet $S$ has the form: \ba \nonumber
S = \frac {g(x,y,t)}{f(x,y,t)} \nonumber \ea
 where
\ba  g &=& {(\bar{f}_{12})}_{0} e^{ - i({\bar{\kappa}}_{2} )^{2}t -
\kappa_{3}y + \bar{\kappa}_{2}x } [1- \frac{ e^{-2\kappa_{1}x}
{(f_{11})}_{0}\,(\kappa_{1} - \bar
\kappa_{2})}{2\kappa_{1}(\kappa_{1} + \bar \kappa_{2})}]\times\nonumber\\
&\times& [1- \frac{ e^{ 2\kappa_{4}y} {(f_{22})}_{0}\,(\kappa_{4} -
\kappa_{3})}{2\kappa_{4}(\kappa_{4} +  \kappa_{3})}], \nonumber\ea
 \ba\nonumber\\
 f &=&  \frac{|(f_{12})_{0}|^{2} \,\,e^{  2 \Re \kappa_{3}y - 2 \Re \kappa_{2}x  - 4 \Re
\kappa_{2} \Im \kappa_{2}t}}{4\Re \kappa_{2}\Re \kappa_{3}} \times
\nonumber\\ &\times&  [1+ \frac{e^{-2\kappa_{1}x}(f_{11})_{0}
|\kappa_{1} - \kappa_{2}|^{2}}{2\kappa_{1}|\kappa_{1} +
\kappa_{2}|^{2}}][1+
 \frac{e^{ 2\kappa_{4}y} (f_{22})_{0} |\kappa_{4} -
\kappa_{3}|^{2}}{2\kappa_{4}|\kappa_{4} +
\kappa_{3}|^{2}}] + \nonumber\\
  &+& [1+ \frac{
e^{-2\kappa_{1}x}\,(f_{11})_{0}}{2\kappa_{1}}] [1+ \frac{
e^{2\kappa_{4}y} \,(f_{22})_{0}}{2\kappa_{4}}],  \ea and one may see
that its asymptotic form at $x, y \rightarrow \infty$ is \be S \sim
\frac{e^{- i \Re\varphi(x, y, t)}}{\cosh(  \Re \kappa_{3}y - \Re
\kappa_{2}x  - 2 \Re \kappa_{2} \Im \kappa_{2}t + \ln
\frac{|c|}{\sqrt{\Re \kappa_{2}\,\Re \kappa_{3}}})},\ee that
coincides with the solution obtained in \cite{Mel'nikov}.

As it has been established earlier, the equations under
consideration admit a "dromion" solution, i.e. a localized solution
that decays exponentially in all directions. In our case if we
impose the conditions \be
\kappa_{1} = \Re \kappa_{2}, \,\,\Im \kappa_{2} = 0,\\
\kappa_{4} = \Re \kappa_{3}, \,\, \Im \kappa_{3} = 0, \ee we obtain
the exponentially localized $N$ - dromion solution for $S$. It looks
like (20), but the block matrices $A$, $B^{n}$, $\alpha $ and
$\tilde{\alpha}$ are $2N \times 2N$, $2N \times 1$ and  $1 \times
2N$ respectively. The $N =1$ - dromion solution has the form : \be S
= \frac {{(f_{12})}_{0} e^{-i (\kappa_{1})^{2}t}\,{\rm
sech}{(\kappa_{1} x + c_{1})} {\rm sech}{(\kappa_{4} y +
c_{4})}}{4[\,\,\zeta + \tanh{(\kappa_{1} x + c_{1})} +
\tanh{(\kappa_{4} y + c_{4})}]}, \ee where the constants are: \ba
\zeta^{2} = \frac{|{(f_{12})}_{0}|^{2} +
{(f_{11})}_{0}{(f_{22})}_{0}}{4\kappa_{1}\kappa_{4}},\nonumber\\
e^{2c_{1}} = \frac{{(f_{22})}_{0}}{2\kappa_{4}(\zeta +
2)\,\zeta},\,\,e^{2c_{4}} = e^{2{c}_{1}}\zeta^{2}.\ea This solution
close to one constructed in \cite{RKLTL}.


\section{Conclusion}

Using the Zakharov - Shabat dressing method we constructed $N$ -
soliton solution for (2+1) - dimensional long wave - short wave
resonance interaction equation (LSRI). We demonstrated that in
generic situation this solution does not decay in $N$ preferred
directions in space infinity. Some choice of constants leads to a
localized solutions that decay exponentially in all directions.

The finding of our study may elicit further investigation. It would
be interesting to analyze in more depth both the solution obtained
for the different choices of constants, and another solutions, as
well as their properties end time evolution.


\begin{thebibliography}{6}
\bibitem[1]{ZS}
V.E.Zakharov, A.B.Shabat, Funct. Anal. Appl. $\bf 8$ 226, 1974.

\bibitem[2]{Mel'nikov}
V.K.Melnikov, JINR-P2-84-13, Jan 1984.

\bibitem[3]{Dodd}
R.K.Dodd, J.C.Eilbeck, J.D.Gibbon, H.C.Morris, "Solitons and
Nonlinear Wave Equations", Academic Press Inc.(London) Ltd., 1984,
and references therein.

\bibitem[4]{ZMNP}
S.P.Novikov, S.V.Manakov, L.P.Pitaevskii, V.E.Zakharov, Theory of
Solitons. The Method of Inverse Scattering (New York:Plenum), 1984.

\bibitem[5]{LC}
Derek W.C.Lai, Kwok W.Chow, Journal of Physical Society of Japan,
$\bf 68$, 6 1847, 1999.

\bibitem[6]{RKLTL}
R.Radha, C.S.Kumar, M.Lakshmanan, X.Y.Tang, S.Y.Lou, Journal of
Physics A: Mathematical and General, $\bf 38$, 44 9649, 2005.

\end{thebibliography}
\end{document}